\documentclass[referee]{raa}            

\usepackage[utf8]{inputenc}

\usepackage{graphicx,times}             
\usepackage{grffile}

\usepackage{natbib}
\usepackage{amssymb,amsmath}
\bibpunct{(}{)}{;}{a}{}{,}
\usepackage{multirow}
\usepackage{ae,aecompl}
\usepackage{fourier}
\usepackage{graphbox}

\usepackage{latexsym}	
\usepackage{gensymb} 
\usepackage{color}

\usepackage{ulem}

\usepackage[pagebackref=true]{hyperref}

\newcommand{\speed}[1]{#1\,km\,s${}^{-1}$}
\newcommand{\uu}{$^{\prime\prime}$}
\begin{document}

\title{Two-sided-loop jet originates from the filament internal reconnection}

   \volnopage{Vol.0 (20xx) No.0, 000--000}      
   \setcounter{page}{1}          

    \author{Yunxue Huang  
      \inst{1,2,*}\footnotetext{$*$Corresponding Authors, these authors contributed equally to this work.}
   \and Jialin Li
      \inst{1,*}
   \and Zhining Qu
      \inst{1}
    \and Ke Yu
      \inst{1}
    \and Hongfei Liang   
      \inst{3}
    \and Rui Xue
      \inst{4}
    \and Xinping Zhou
      \inst{1}
   }

   \institute{College of Physics and Electronic Engineering, Sichuan Normal University, Chengdu 610068, China; {\it xpzhou@sicnu.edu.cn}\\
        \and
             State Key Laboratory of Solar Activity and Space Weather, Beijing 100190\\
        \and
             Department of Physics, Yunnan Normal University, Kunming 650500, China\\
        \and
		Department of Physics, Zhejiang Normal University, Jinhua 321004, China ;{\it ruixue@zjnu.edu.cn }\\
   \vs\no
   {\small Received 20xx month day; accepted 20xx month day}}

 \abstract{ Magnetic reconnection driving two-sided-loop jet is typically associated with interactions between an emerging bipole and the overlying horizontal magnetic field, or between filaments from separate magnetic systems. Leveraging high temporal and spatial resolution observations from ground-based and space-borne instruments, we have identified a two-sided-loop jet originating from magnetic reconnection between threads within a single filament. Our observations show that as two initially crossing filamentary threads within the filament converge, reconnection takes place at their intersection. In the Doppler images, distinct redshift and blueshift signals are observed at the locations where the filament threads intersected. This process generates a two-sided-loop jet with outflow speeds of \speed{22.2} and \speed{62.5}. Following reconnection, the original crossing threads transform into two parallel threads that subsequently separate at speeds of \speed{2.8} and \speed{8.3}. This observation offers a new perspective on the mechanisms responsible for jet formation.
}

   \authorrunning{Yunxue Huang \& Jialin Li et al. }            
   \titlerunning{two-sided-loop jet originates from the filament internal reconnection}

   \maketitle


\section{Introduction}
\label{sec:intro}
Solar jet is a stream of plasma that moves along large-scale open collimated magnetic field lines or far-reaching coronal loops, with a width range of $10^2-10^5$\,km \citep{2000ApJ...542.1100S}. On morphology, the solar jets can be classified into two types: the anemone jets and the two-sided-loop jets, which are traditionally explained by the emerging-flux model \citep{1994xspy.conf...29S,1995Natur.375...42Y,1996PASJ...48..123S,1998SoPh..178..379S}. As one of the basic explosive activities in the solar atmosphere, the research on it can help us to learn more about the magnetic reconnection process.

The chromospheric anemone-like jets are those straight plasma beams with an observation characteristic of a bright cusp and an inverted Y-shape structure, with a typical length of $1.0-4.0$ Mm, a width of $100-400$ km, a velocity of \speed{$5-20$} and a lifetime of $100-500$\,s \citep{2011ApJ...731...43N}. While the two-sided-loop jet is a pair of plasma beams ejected in opposite directions from the source region. Coronal jets typically exhibit an apparent velocity of about \speed{200} \citep{2016SSRv..201....1R}. Their typical lifetimes are 20 (30) minutes in the 171 (304) \AA\ \citep{2009SoPh..259...87N}. Rotating and transverse motions are two typical characteristics of solar jets \citep{2007PASJ...59S.771S,2012RAA....12..573C,2013ApJ...769..134M,2024SCPMA..6759611Z,2024ApJ...974L...3Z}. Sometimes both the hot and cold components can coexist in the jet \citep{2013ApJ...775..132J,2022ApJ...936...51W}, the cold matter of the blowout jet originates from the erupting filaments in the eruption source region  \citep{2012ApJ...745..164S,2017ApJ...851...67S}. Specifically, \citet{2019ApJ...883..104S} observed two different temperature components in the two-sided-loop jet, and proposed that the cold plasma was from the mini-filament eruption under the large filament. \citet{1995Natur.375...42Y,1996PASJ...48..353Y} first simulated the anemone jet and two-sided-loop jet without considering the effect of heat conduction and radiative cooling. \citet{2013ApJ...777...16Y} generated repeated high-temperature and high-density plasmoids in the simulation, and these plasmoids ejected upward and downward simultaneously, the upward moving plasmoids corresponded to the anemone jet.

As the two-sided-loop jet is relatively rarely observed, it makes people's research on it not very mature, especially on the triggering mechanism there is no unified theory. Though we have known that the fundamental and necessary physical process is magnetic reconnection \citep{2000mare.book.....P,2022LRSP...19....1P}, the component joining in the reconnection, the position of the reconnection site and driving source are various.
\citet{1992PASJ...44..265S,1994xspy.conf...29S} proposed the emerging-reconnection model: when the emerging magnetic flux reconnects with the open vertical or oblique field, this produces an anemone jet; while the reconnection is with the overlying horizontal coronal magnetic field, a two-sided-loop jet is formed. These models are supported by many observations \citep[e.g.,][]{2013ApJ...771...20M,2014SSRv..186..227S,2016SSRv..201....1R,2018ApJ...854..174T}. However, currently the vast majority of studies indicate that magnetic flux cancellation, rather than magnetic flux emergence, is the primary cause of most solar jets \citep[e.g.,][]{2008A&A...478..907C,2012RAA....12..300Y,2013ApJ...764...70Z,2014ApJ...783...11A,2015ApJ...814L..13L,2016ApJ...832L...7P,2017ApJ...835...35H,2017ApJ...842L..20L,2021RSPSA.47700217S,2024ApJ...964....7Y}. Besides, many jet phenomena are related to the activities of filaments. Solar jets can provide material accumulation for the formation of filaments and even cause filament instability, leading to filament oscillations \citep[e.g.,][]{2014ApJ...785...79L,2017ApJ...851...47Z,2019ApJ...872..109A}. The generation of solar jet (including the two-sided-loop jet) is often related to the reconnection associated with filaments or filament eruptions. \citet{2019ApJ...883..104S} reported a two-sided-loop jet associated with the eruption of a mini-filament located below a large overlying filament. The eruption had two stages, and an external and an internal magnetic reconnection process were observed respectively. It is very likely that the two-sided-loop jet in this event was produced by the alternating occurrence of magnetic flux cancellation and emergence. \citet{2019ApJ...871..220S} reported a two-sided-loop jet event caused by the eruption of a small-scale filament. The magnetic triggering mechanism of this jet is still magnetic cancellation. \citet{2017ApJ...845...94T} reported a two-sided-loop jet driven by the reconnection between two adjacent filamentary threads. Additionally, \citet{2016ApJ...816...41Y} also found a similar phenomenon in an event in which filament formed by the reconnection of the threads of two different filaments. 

Whether the emerging-reconnection model or the filaments reconnection model, the reconnection occurs always between two magnetic systems, while in this letter, we will report a two-sided-loop jet caused by an internal reconnection of the magnetic system itself. This letter will be developed in the following order:  Section \ref{sec:Obs} briefly introduces the observation instruments; Section \ref{sec:Results} gives the observational analysis and results; Section \ref{sec:discussion} presents the discussion and conclusion.

\section{Observations}
\label{sec:Obs}

The two-sided-loop jet event, occurring on November 19, 2019 in active region (AR) NOAA 12752, was simultaneously observed by the New Vacuum Solar Telescope \citep[NVST;][]{2014RAA....14..705L,2020ScChE..63.1656Y}, Atmospheric Imaging Assembly \cite[AIA;][]{2012SoPh..275...17L} and Helioseismic and Magnetic Imager \cite[HMI;][]{2012SoPh..275..229S} on board the Solar Dynamics Observatory \citep[SDO;][]{2012SoPh..275....3P}. NVST's spatial resolution is 0\uu.3 in both line center (6562.8 \AA) and line wings (6562.8$\pm$1 \AA) images, and its minimum temporal cadence is 12\,s. If the observation of line wings is requested, the cadence of H$\alpha$ observations is scaled down to about 50\,s, and in such a case, line wings' snapshot share the same cadence with H$\alpha$ snapshots. The NVST images are co-aligned with each other by using a high accuracy solar image registration procedure \citep{2015RAA....15..569Y}. The time resolution and pixel size of the AIA images are 12\,s and 0\uu.6, respectively, which are calibrated using the standard program \texttt{aia\_prep.pro} in the solar software. The cadence and pixel size of the line-of-sight magnetograms taken by the HMI and measuring precision are 45\,s, 0\uu.5, respectively. Here, we mainly use the AIA 171 \AA, 193\AA, 211\AA\ and 304 \AA\ as well as the NVST line center (6562.8 \AA) and line wings (6562.8$\pm$1 \AA) images to study this event. The images of these two instruments were co-aligned with each other by carrying out an automatic mapping approach developed by \citet{ji}.

\section{Results}
\label{sec:Results}

\begin{figure}
	\centering
	\includegraphics[width=1\linewidth]{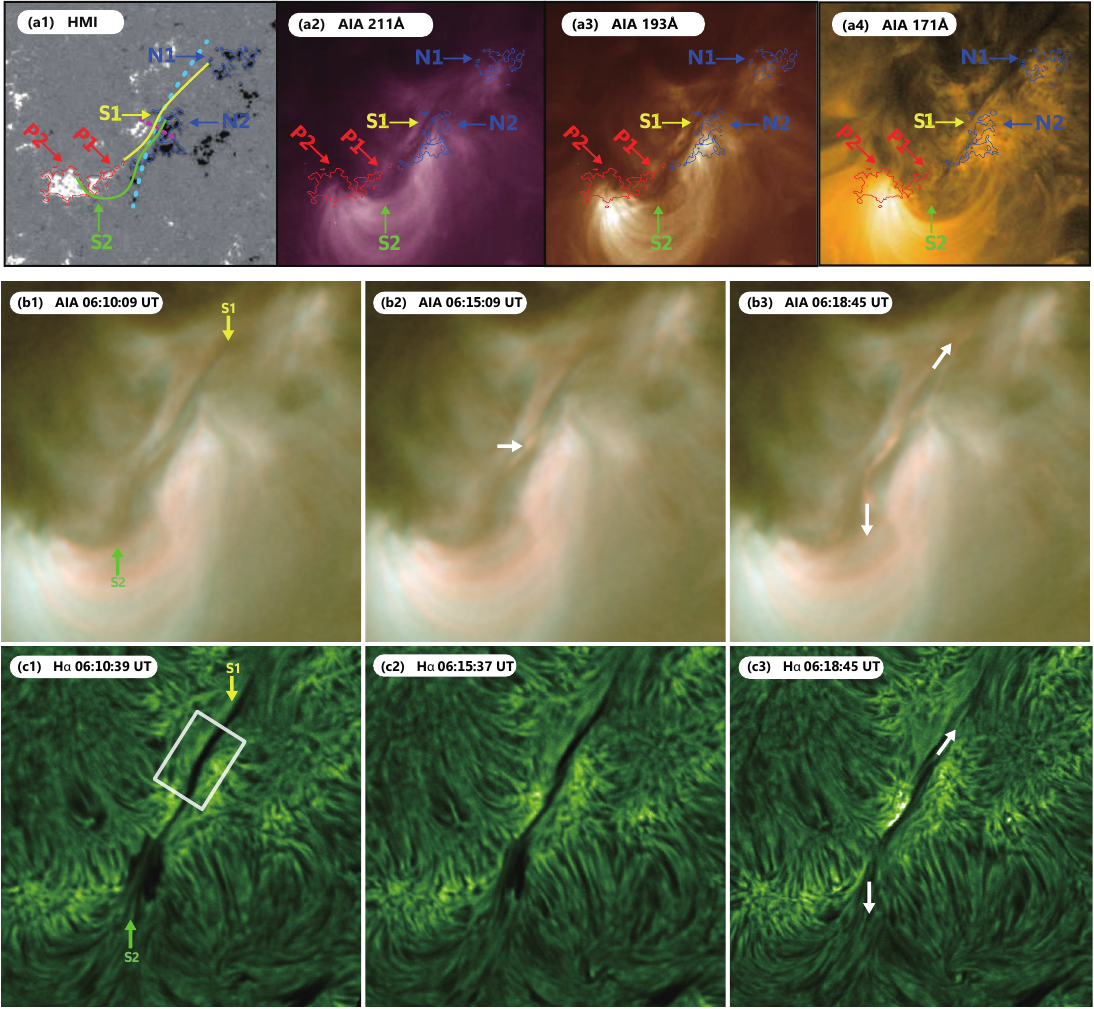}
	\caption{The four panels (a1)–(a4) are close-up HMI, AIA 211 \AA\, 193 \AA\, 171\AA\, respectively. In panels (a1)–(a4), the two threads are marked by S1 and S2, respectively. The red and blue contours in panels (a1)–(a4) are the footpoints of two threads at the level of ±50 Gauss, respectively, which are labeled with P1, P2, N1 and N2 (N represents the negative polarity and P the positive). Panels (b1)-(b3) and (c1)-(c3) are AIA composited snapshots, H$\alpha$ center snapshots, respectively. The two threads are labeled with S1 and S2 in panel (b1) and (c1), respectively. The white arrow in panel (b2) symbolizes the brightening occurring in-between two threads. The arrows in (b3) and (c3) the extending of the two-sided-loop jet.}
		\label{fig:overview}
\end{figure}

\ref{fig:overview} (a1)-(a4) give the overview of the pre-eruption of this event (at about 06:10 UT). \ref{fig:overview} (a1)–(a4) are HMI, AIA 211 \AA\, 193 \AA\, 171\AA\ snapshots, in which the field-of-view (FOV) of four close-up snapshots all are 120\uu\ by 120\uu. We outline two threads of one filament with the yellow and green curves marked by "S1" and "S2" in \ref{fig:overview} (a1), whose footpoints (N1, N2, P1 and P2; "N" labeling negative polarities and "P" positive polarities) are contoured by blue and red profiles at the level of ±50 Gauss (\ref{fig:overview} (a2)–(a4)); the blue for negative polarities and the red for positive polarities. S1 was rooted in N1P1 and S2 in N2P2. The filamentary threads' configuration can be seen roughly from \ref{fig:overview} (a2)–(a4): they were crossed and slightly separated from each other, in an overall shape of J. We can diagnose the evolution of the two-sided-loop jet through multi-wavelength observations. The results are displayed in \ref{fig:overview} (b1)–(c3), in which the two rows are the composited AIA snapshots (171\AA, 193\AA\ and
211\AA), H$\alpha$ snapshots; their FOVs all are 100\uu\ by 100\uu. Composited AIA snapshots are made through superposing three semitransparent AIA snapshots. During the erupting onset (at about 06:10 UT), two threads seemed like remaining stable, characterized by no brightening or other obvious activities at all in EUV channels and also H$\alpha$ line core (\ref{fig:overview} (b1) and (c1)). A few minutes latter, at about 06:15 UT, the part in-between two threads began to brighten (see the white arrow in \ref{fig:overview} (b2)), and the separated and crossed features of these two threads can be distinguished in EUV channels and H$\alpha$ channel (see the boxes in \ref{fig:overview} (b2) and (c2). And then, the outflows escaped along two threads during next several minutes, forming two bright spires directed oppositely, which consisted of a two-sided-loop jet (see the white arrows in \ref{fig:overview} (b3)).

\begin{figure}
	\centering
	\includegraphics[width=1\linewidth]{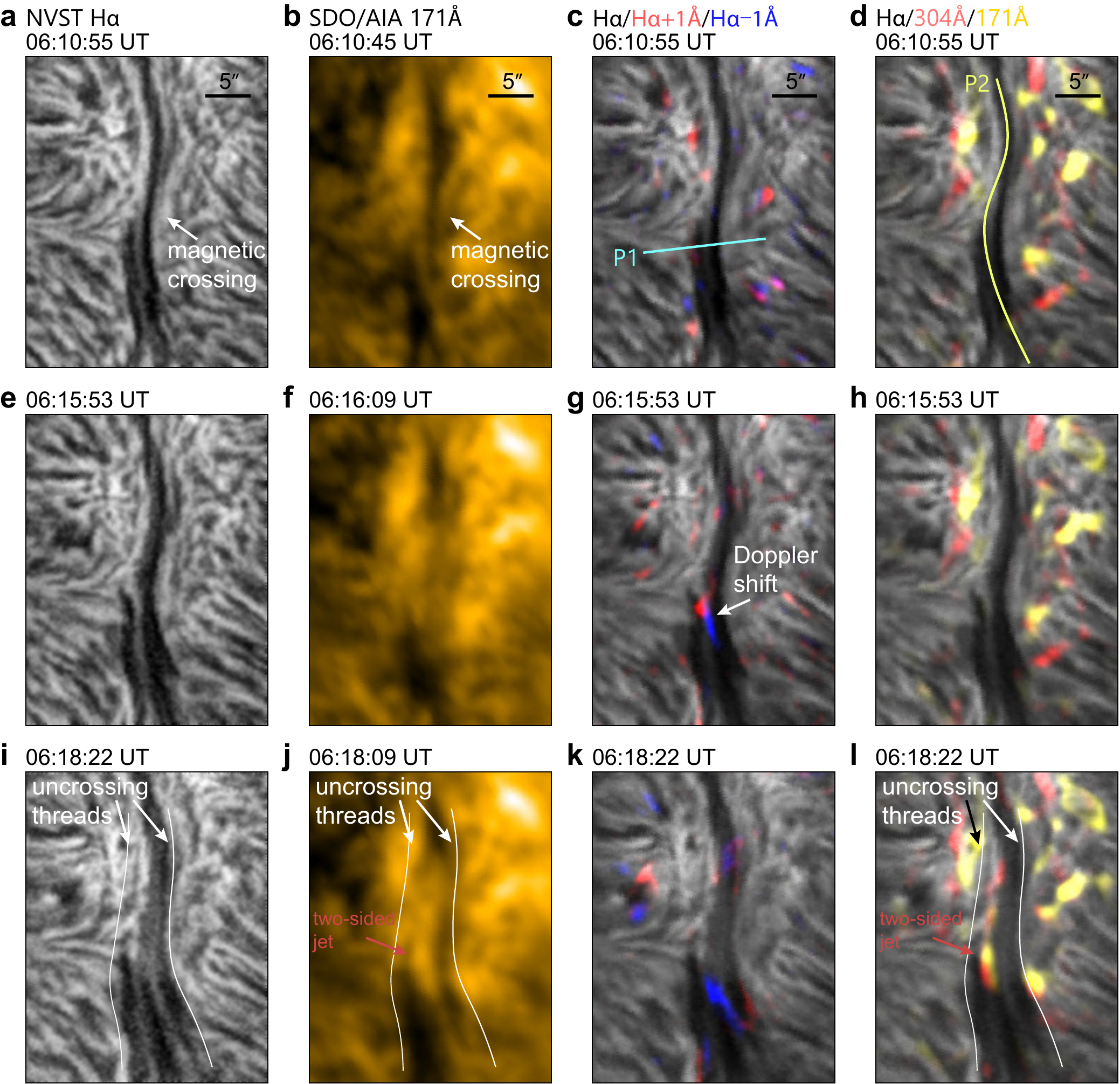}
	\caption{The four columns of \ref{fig:evolution} are NVST H$\alpha$  images, AIA 171 \AA\ images, composite images of H$\alpha$ and the Doppler images composed of H$\alpha$-wing observation, composite images of AIA 171 \AA\ and 304 \AA\, respectively. The time range is from 06:10 UT to 06:18 UT. Panels (a)-(d) show the alignment of the two crossing filament threads prior to the jet.The white arrows in panels (a)-(b) represent the alternating points of the two threads. The line labeled P1 is at the interleaved point and is the slice position of time-distance diagram of inflow and outflow shown in \ref{fig:kinematic}. Line P2 refers the position of the slice along the direction of the two jets and is used to study the velocity of the two jets shown in \ref{fig:kinematic}. Panels (e)-(h) show the reconnection process, and the white arrow in panel (g) indicates the observed Doppler red-blue shift pairs. Panels (i)-(l) are the image after the reconnection is completed and shows the initial jet. Two white lines are used in panels (i)(j)(l) to trace the two threads parallel to each other after the reconnection, and red arrows are used in panels (j) and (l) to indicate the newly appeared jet.}
		\label{fig:evolution}
\end{figure}

\ref{fig:evolution} displays the evolution process of the two threads within the filament through the further magnified NVST H$\alpha$ line center images, raw AIA 171 \AA\ images, composite H$\alpha$ line center images and H$\alpha$ line-wing observations, composite images of H$\alpha$, AIA 171 \AA\ and 304 \AA. As shown in \ref{fig:evolution} (a)-(d), we can observe a north-south oriented filament with a width of 2\uu-5\uu, consisting of two threads crossing in central region indicated by white arrows in \ref{fig:evolution} (a) and (b). At 06:15 UT, these two threads began to move close to each other (see \ref{fig:evolution} (e)-(h)), which was also reflected in Doppler images: we can observe signals of both redshift and blueshift at their overlapping and intersecting location (see the white arrow in \ref{fig:evolution} (g)). Subsequently, a significant brightening appeared at the crossing point, which indicated that the reconnection occurred near the crossing point. Please refer to the animation of \ref{fig:evolution} for the detailed interaction process. Several minutes later at about 06:18 UT, in the H$\alpha$ observation images, we can detect that the two original crossing threads formed two new parallel threads (see \ref{fig:evolution} (i)-(l)). This change in position indicated that the reconnection of the two threads had been completed. This process can be identified for the  AIA 171 \AA\ images: centered at the interlaced reconnection point, a bright hot plasma stream was observed moving in opposite directions along the path of the two threads' extension (indicated by the red arrow in \ref{fig:evolution} (j)). The two-sided-loop jet also has a clear response in the composite image of H$\alpha$ AIA 171 \AA\ and 304 \AA\, as shown in \ref{fig:evolution} (l), revealing a clear bidirectional material flow between the newly formed parallel threads. In the composite image of H$\alpha$-wing, as shown in \ref{fig:evolution} (k), it further proved that reconnection had completed at this time. Additionally, the red and blue shift in the direction of outflow can be seen in the extension direction of the two threads, indicating that there was material ascending and descending at this time.

\begin{figure}
	\centering
	\includegraphics[width=.6\linewidth]{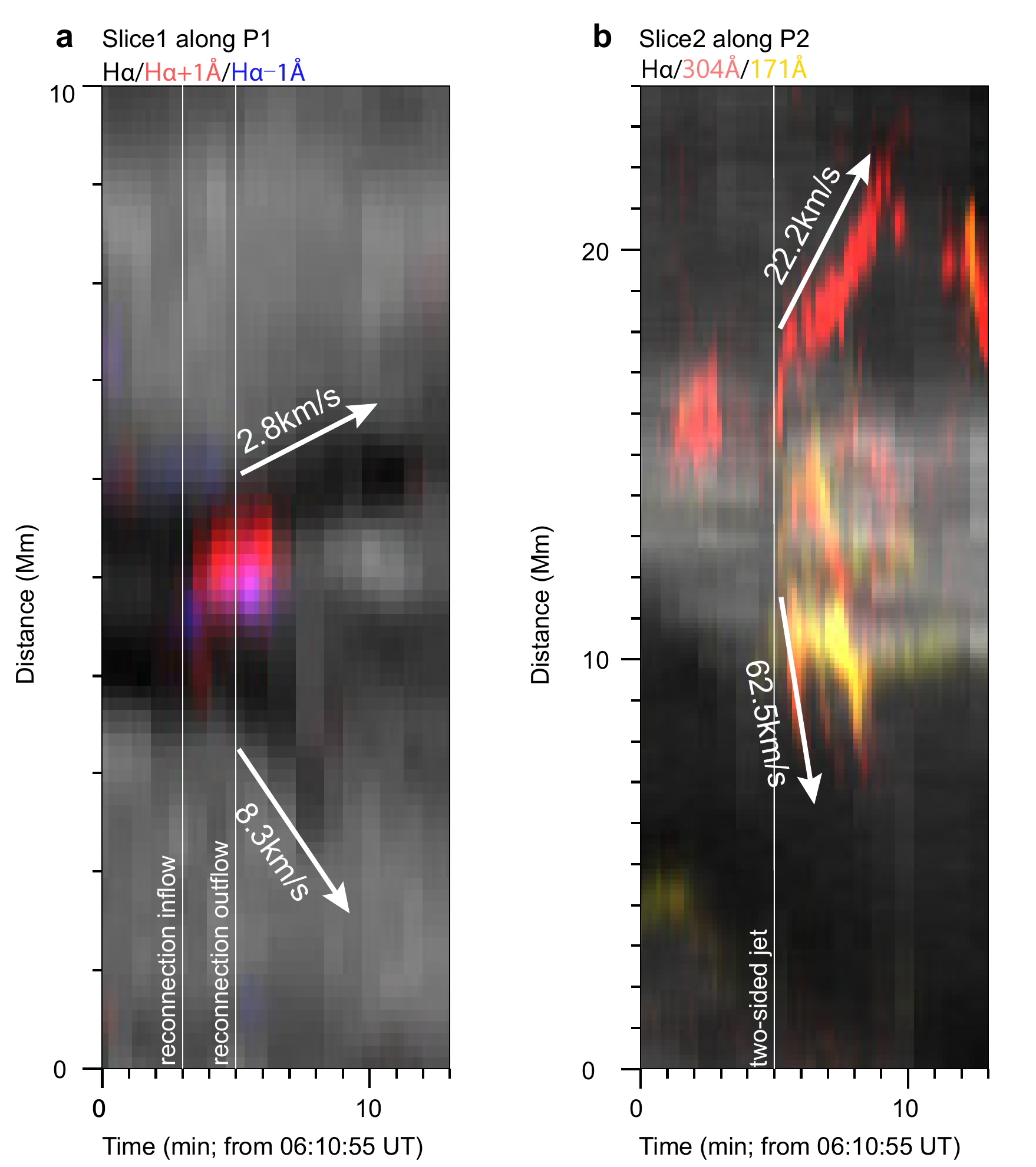}
	\caption{Panel (a) is made along line P1 in \ref{fig:evolution} panel (c). The two vertical lines correspond to the start time of inflows and outflows respectively. Two white arrows indicate the direction of inflow and outflow, with the speeds of \speed{2.8} and \speed{8.3}, respectively. Panel (b) is made along line P2 in \ref{fig:evolution} panel (d), showing the velocity and start time of the jet in both directions. The speeds of the two branches are \speed{22.2} and \speed{62.5}, respectively. }
		\label{fig:kinematic}
\end{figure}

\begin{figure}
	\centering
	\includegraphics[width=1\linewidth]{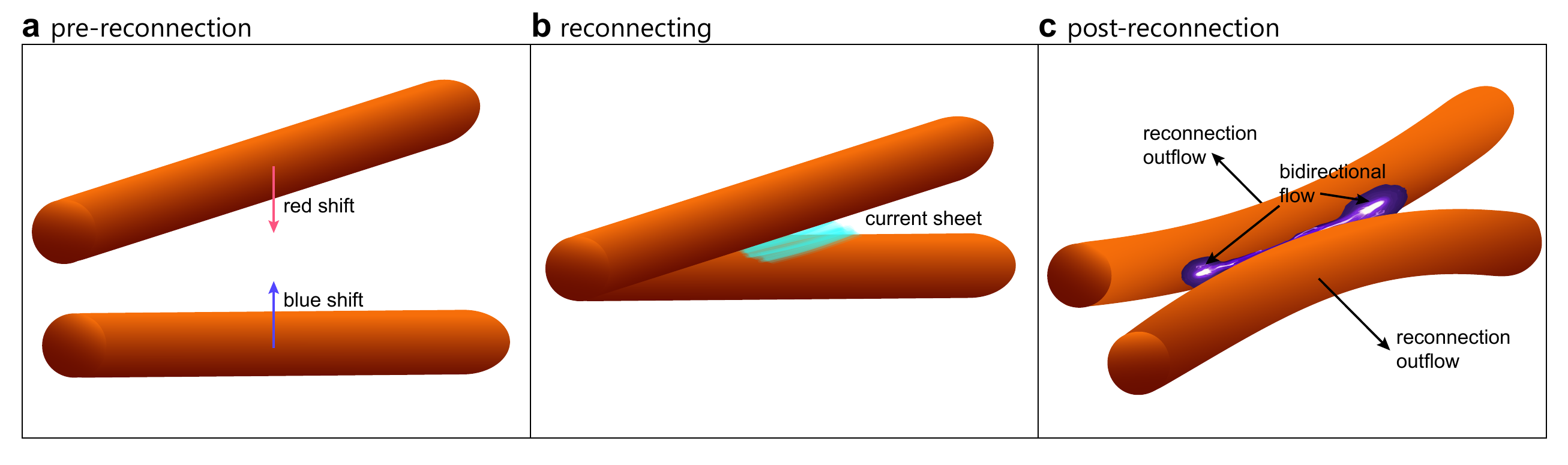}
	\caption{Panels (a)-(c) represent the three stages pre-reconnection, reconnecting and post-reconnection respectively. Panel (a) represents the red and blue shift pairs of the pre-reconnection stage, panel (b) represents the current sheet formed at the crossing point in the reconnection, and panel (c) represents the outflow along the two directions at the end of the reconnection and the two-sided-loop jet formed between them.}
		\label{fig:cartoon}
\end{figure}

\begin{figure}
	\centering
	\includegraphics[width=1\linewidth]{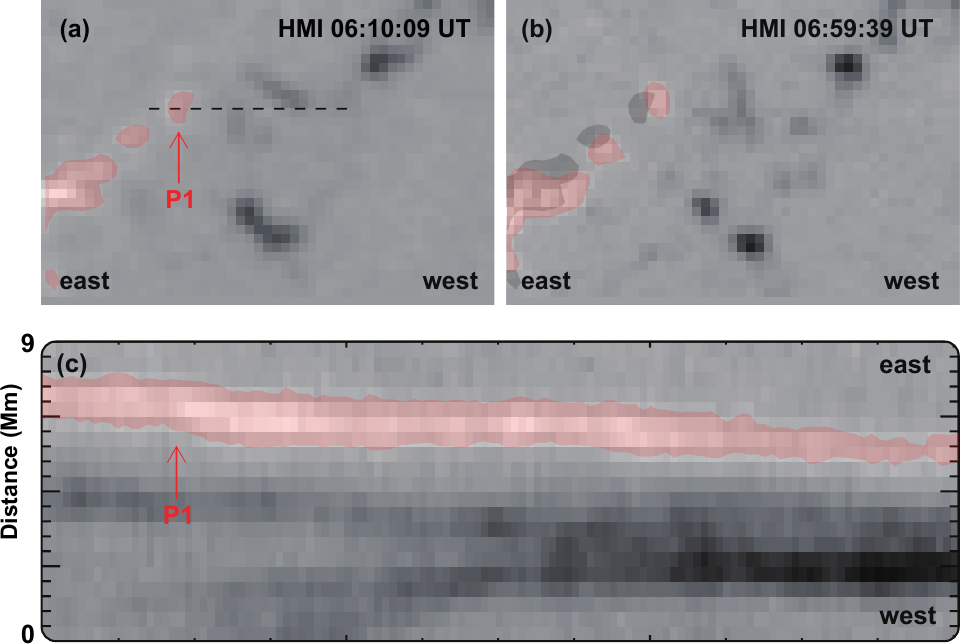}
	\caption{Analysis for the trigger of reconnection. Panels (a) and (b) show the location of P1 at two different moments (06:10:09 UT and 06:59:39 UT, respectively), indicated by red shielded regions. For the comparison, the location of P1 at 06:10:09 UT is contoured into panel (b) with the gray shielded regions. Panel (c) is time-distance stack plot, whose paths are along the black dash line in \ref{fig:HMI} (a). The red shield region in panel (c) indicates the contour of P1. }
		\label{fig:HMI}
\end{figure}

To analyze the jet's evolution in detail, time-distance plots are generated from the composite images and are presented in \ref{fig:kinematic}. Panels (a) and (b) of this figure display time-distance along slits P1 and P2 (see \ref{fig:evolution}). Slit P1 is a straight line perpendicular to the filament near the crossing point. Slit P2 is a curved line segment along the filament thread, directing from north to south.  \ref{fig:kinematic} (a) indicates that reconnection occurred around 06:13:55 UT. Subsequently, at approximately 06:15:55 UT, the filament exhibited significant outflow motion. As shown in \ref{fig:kinematic} (a), the outflow velocity was approximately \speed{2.8} eastward and \speed{8.3} westward. Furthermore, significant redshift and blueshift signals (indicated by red and blue regions in \ref{fig:kinematic} (a)) were observed prior to the outflow onset. This suggests that the mutual approach of the threads led to magnetic reconnection at their intersection, which in turn generated outflows. Almost immediately following the outflow's appearance, two-sided-loop jet also began to form, as depicted in \ref{fig:kinematic} (b). The ejections along the two branches of the jet spire showed notable differences. As illustrated in \ref{fig:kinematic} (b), the southward ejections were generally more vigorous, ejecting a larger amount of plasma. Moreover, the southward ejections had appeared periodically at least three times, with a slightly higher velocity (\speed{62.5}) compared to the northward ejections (\speed{22.2}). The northward ejections mainly manifested in the AIA 304 \AA\ observations, while the southward ejections mainly appeared in the AIA 171 \AA\ observation. This result may imply that the kinetic and thermal energy released by the reconnection was relatively less in the northward ejections compared with the southward ejections. This asymmetry might be due to the projection effect, where the apparent velocity is merely a component of the true velocity. Additionally, it is not difficult to find that the integral velocity of the jet is lower than those of most other jets reported in previous studies \citep{1999SoPh..190..167A,2013ApJ...775..132J,2017ApJ...841L..13C,2018ApJ...861..108Z,2019ApJ...887..220Y,2023A&A...678L...7W}. It is almost the lower limit of the typical jet velocity. One possible reason is that the local Alfvén speed is relatively low, which means that the comparatively weak magnetic field and high plasma density may lead to a constricted acceleration proccess of the jet.

\section{Discussion and conclusion}
\label{sec:discussion}

Combined with the high spatial and temporal resolution of ground-based telescope NVST and the space telescope SDO, we presented a two-sided-loop jet which occurred on 2019 November 19 in active region 12572. By diagnosing the formation mechanism of the two-sided-loop jet, we find that the evolution process of the filaments associated with the jet may be caused by the reconnection between two crossing threads of a filament. The multiple observations revealed that the two intersecting threads of a filament approached each other, and magnetic reconnection occurred at their intersection, thus leading to the generation of the two-sided-loop jet. This observation sheds a new light to the formation of the two-sided-loop jet: two-sided-loop jet can also be produced by the internal reconnection between two threads of a filament.

Solar jets are prevalent throughout the solar atmosphere and are frequently observed in association with magnetic flux emergence and cancellation activities. Theoretical models propose that the two-sieded loop jet can resulted from the emerging magnetic flux reconnects with the overlying horizontal coronal magnetic field \citep{1994xspy.conf...29S,1995Natur.375...42Y,2019ApJ...887..220Y}. Observational evidence supports this model, with some cases demonstrating two-sided-loop jet production resulting from magnetic flux emerging from below the photosphere and reconnecting with a pre-existing overlying horizontal field \citep{1998SoPh..178..173K,2022MNRAS.516L..12T}. Previous studies have revealed that there is a close relationship between bidirectional jets and filaments. For example, the eruptions of mini-filaments or magnetic reconnections associated with mini-filaments can trigger bidirectional jets \citep{2019ApJ...871..220S,2019ApJ...883..104S,2019ApJ...887..220Y}. \cite{2017ApJ...845...94T} studied a two-sided-loop jet which was induced by reconnection between two filaments approaching each other. These observations collectively demonstrate that the occurrence of two-sided-loop jets could result from the external reconnection between the two different filaments. \cite{2020ApJ...899...19C} proposed a possible mechanism for coronal mini-jets occurring in tornado-like prominences. Considering that the magnetic field in this situation might be highly complex, these mini-jets could be caused by internal reconnection between adjacent threads. This mechanism offers a compelling explanation for the fine structural features and ejection directions of min-jets. \cite{2023ApJ...958..116W} found that during the activation process of a filament, once the filament became unstable, it triggered internal reconnection between different magnetic field lines that supported the magnetic dips, resulting in a series of localized brightenings within the filament. Subsequently, external reconnection occurred between the filament threads and the surrounding magnetic field, leading to a unidirectional flow towards the eastern footpoint of the filament. However, none of these internal reconnections gave rise to a direct bidirectional jet.

\citet{2021NatAs...5...54A} demonstrated that the nanojet is a consequence of the slingshot effect from the magnetically tensed, curved misaligned magnetic field lines reconnecting at small angles in numerical experiment and observation. However, they did not observe the significant two-sided-loop jet due to the reconnection. \citet{2024A&A...692A.119C} reported a discovery of small-scale, elongated brightenings propagating bidirectionally along threads within arch filament systems, observed by Solar Orbiter/EUI. These brightenings, characterized by speeds of \speed{$100-150$} and dominated by cool plasma emission, are attributed to magnetic reconnection triggered by newly emerging magnetic flux interacting with pre-existing magnetic fields within the arch filament systems. The observation presented in this paper provides an alternative new triggering mechanism for explaining two-sided-loop jets: the two-sided-loop jet could be triggered by the reconnection within a filament. Based on the analysis, we present a simple cartoon model in \ref{fig:evolution} to vividly illustrate the formation and evolution process of the two-sided-loop jet. The topological structure before the ejections is shown in \ref{fig:evolution} (a), where the two threads within the filament intersect at their central region, forming an X-shape structure. As they approach each other, they exhibit redshift and blueshift in Doppler images, and ultimately, reconnection takes place at the intersection as shown in \ref{fig:evolution} (b). Subsequently, the outflow and directional flow occur due to the reconnection. Consequently, the originally intersecting threads were transformed into newly formed parallel threads as shown in \ref{fig:evolution} (c), where the plasma ejection along the direction of the newly formed threads is shown in purple.

To rule out the possibility of the magnetic emergence or cancellation mechanism, we conducted coordinated observations using HMI. \ref{fig:HMI} (a) and (b) are two HMI snapshots at two different moments, while (c) is the time-distance stack plots. Here, \ref{fig:HMI} (c) was made by composing the time sequence of the intensity profiles of HMI images along the black dash line in \ref{fig:HMI} (a). The location of footpoint P1 at 06:10:09 UT and 06:59:39 UT (before and after eruption) is contoured by red shielded region (50 Gauss) in \ref{fig:HMI} (a) and (b), respectively. For comparing the location of P1 at two different moments, we also contour P1 at 06:10:09 UT onto \ref{fig:HMI} (b) but mark it by the gray shield region. From the comparison of P1's location (\ref{fig:HMI} (b)), we can find its slow western motion. Such scenario can also be found in \ref{fig:HMI} (c) (see the red contoured region), in which P1 shows its relatively obvious western motion at about 06:15 UT. There is no signature of magnetic flux emergence or cancellation. On the contrary, it is the motion of one footpiont of threads that resulted in the contact between two threads, and further triggered the reconnection among these two threads at their crossed site.

The event reported in this paper provides observational evidence that bidirectional jets can be driven by magnetic reconnection between adjacent filament threads within a filament system. To date, such observations remain scarce, and additional case studies combined with theoretical modeling are required to elucidate the detailed physical mechanisms governing these dynamics. As the Sun approaches the maximum of solar cycle 25 \citep{2023SCPMA..6629631C}, this phase of heightened activity presents a unique opportunity to leverage synergetic observations from the SDO and recently launched Chinese solar missions. These include the Advanced Space-based Solar Observatory  \citep[ASO-S;][]{2019RAA....19..156G,2023SoPh..298...68G}, the Chinese H$\alpha$ Solar Explorer  \citep[CHASE;][]{2022SCPMA..6589602L}, and the Solar Upper Transition Region Imager \citep[SUTRI;][]{2017RAA....17..110T}. The multi-instrument, multi-wavelength capabilities of these observatories will enable unprecedented investigations into the initiation, propagation, and thermodynamic properties of bidirectional jets in filament systems.

\begin{acknowledgements}
We are thankful to the teams of SDO and NVST for providing excellent data. This work is supported by the Natural Science Foundation of China (12303062, 12203043), the Sichuan Science and Technology Program (2023NSFSC1351,  2025ZNSFSC0315), Key Laboratory of Detection and Application of Space Effect in Southwest Sichuan at Leshan Normal University, Education Department of Sichuan Province(No.ZDXM20241002)), and Project Supported by the Specialized Research Fund for State Key Laboratory of Solar Activity and Space Weather.  We gratefully acknowledge Sichuan Normal University Astrophysical Laboratory Supercomputer for providing providing the computational resources. 
\end{acknowledgements}


\label{lastpage}

\end{document}